\documentclass[prl,twocolumn,10pt]{revtex4}
\usepackage{graphicx}
\usepackage{feynarts}
\usepackage{longtable}

\def\text#1{\mathrm{#1}}

\begin{document}
\title{The two-loop QCD amplitude $gg \to h,H$  
in the Minimal Supersymmetric Standard Model}
\author{Charalampos Anastasiou}
\email{babis@itp.phys.ethz.ch}
\author{Stefan Beerli}
\email{sbeerli@itp.phys.ethz.ch}
\author{Alejandro Daleo}
\email{adaleo@itp.phys.ethz.ch}
\affiliation{Institute for Theoretical Physics, ETH, CH-8093, Z\"urich, 
Switzerland}
%\date{\today}
\pacs{11.10.Gh,11.30.Pb,12.38.Bx,12.60.Jv,14.80.Cp}
\begin{abstract}
We present the two-loop QCD amplitude for 
the interaction of two gluons and a CP-even Higgs
boson in the Minimal Supersymmetric Standard Model.
We apply a novel numerical method for the evaluation 
of Feynman diagrams with  infrared, ultraviolet and 
threshold singularities. 
We discuss subtleties in the ultraviolet renormalization 
of the amplitude with conventional dimensional regularization, 
dimensional reduction, and the four dimensional helicity scheme.
Finally, we show numerical results for scenarios of supersymmetry breaking 
with a rather challenging phenomenology in which the Higgs signal in the 
MSSM is suppressed in comparison to the Standard Model. 

\end{abstract}
\maketitle

The loop-mediated interaction of a Higgs boson and gluons 
is  the main  production mechanism for a Higgs boson at hadron 
colliders.  Depending on the decay channels in which 
a Standard Model Higgs boson may be discovered, 
the measurement of the signal 
cross-sections can be achieved with a precision of about 
$\pm 10 \%$ or better~\cite{cmstdr}. Amid the discovery of a Higgs 
boson, the signal cross-section will be an independent 
precision test of the Standard Model and its extensions. 

The gluon-fusion Higgs boson production cross-section is sensitive 
to higher order QCD  
corrections~\cite{Tspira,smeffective,Vharlander}. 
At the LHC, the signal cross-sections are known to change up to a 
factor of two when perturbative corrections from  
${\cal O }\left(\alpha_s^2\right)$  through  
${\cal O }\left(\alpha_s^4\right)$ are 
included~\cite{Fehip,WWnnlo}. 
The accuracy of these theoretical predictions is about 
$\pm 10 \%$. The partonic decay width to gluons also 
increases by a factor of two when including the known 
higher order QCD corrections. This quantity is now computed through 
order ${\cal O }\left(\alpha_s^5\right)$ with an accuracy better than
 $1\%$~\cite{baikov}.  
Theoretical uncertainties in the $ggh$ interaction within 
the Standard Model  are currently  adequately small for a 
future comparison with LHC data at a  $10 \%$ precision level.

Extensions of the Standard Model (SM) postulate diverse mechanisms 
for breaking the electroweak symmetry.
A different Higgs boson sector than the one in the Standard Model 
and additional new particles are often introduced.  
The effects of undiscovered particles on the gluon-fusion cross-section 
are rather unconstrained from experimental data. 
Novel colored particles can change the Higgs 
and gluon interaction dramatically. For example, an additional 
heavy-quark with SM-like Yukawa coupling to the Higgs boson 
would contribute almost as much as the top-quark to the  $ggh$ amplitude. 
Recent examples in well motivated models were  shown 
in~\cite{falkowski}.

The complexity of the two-loop SM computations 
at ${\cal O }\left(\alpha_s^3\right)$  in the full theory  
and at ${\cal O }\left(\alpha_s^4\right)$  in the limit 
of a heavy top-quark is serious.  
The methods that have been employed are powerful 
and may be employed in other models.  
In especially simple modifications of the 
SM Lagrangian, for example adding a fourth generation with heavy leptons 
and quarks, the existing calculations are already sufficient. 
However, it will be important to know the gluon-fusion cross-section 
through at least order ${\cal O} \left(\alpha_s^3\right)$ in all models 
which aspire to explain LHC data. 
Many viable extensions of the SM contain colored 
particles where, at two-loops, more than one of  these massive 
particles appears in diagrams contributing to 
the gluon-gluon-higgs amplitude.  
Known analytic methods for two-loop calculations 
are restricted to problems with a small number of mass parameters. 
New techniques to evaluate multi-loop integrals with diverse 
mass-scales are indeed required. 

In this article, we compute the two-loop 
QCD amplitude $gg\to h,H$ for a CP-even light $(h )$ and heavy $(H)$ 
Higgs boson  in the minimal supersymmetric extension of the Standard Model 
(MSSM). We employ a numerical method which we have recently developed 
for multi-loop calculations~\cite{beerli,lazopoulos}. 
In our method, Feynman diagrams with diverse 
combinations of massive and massless propagators are treated on an 
equal footing. Given the complexity of the MSSM, the computation of 
$2 \to 1$ two-loop amplitudes in other extensions of the SM may also 
be tractable. 

Partial SM-like contributions from supersymmetric 
diagrams with only squarks in the loops have been recently 
computed  in the literature~\cite{bucherer,spira-squarks,bonciani1,bonciani2}. 
The MSSM amplitude is also  
known in the limit of a light Higgs boson with respect to quarks, 
squarks, and gluinos~\cite{mssmeff1,mssmeff2,mssmeff3}. 

The computation of the two-loop amplitude 
without using an effective theory approach is well motivated 
in SM extensions. Relatively light colored particles 
(lighter than the the top-quark) are not excluded experimentally.  
Heavy Higgs bosons are also  predicted in the spectrum of new theories.  
Also, the couplings of the bottom quark to  Higgs bosons  
may be  significantly enhanced in models with more than one 
Higgs  doublet. 
The MSSM exhibits all of these features; 
knowledge of the amplitude for the gluon-gluon-Higgs 
interactions without assumptions about the mass hierarchy of Higgs 
bosons and colored particles is therefore especially important.  
We consider the  MSSM  an archetype for many other models regarding 
its computational challenges. 
To the best of our  knowledge, we present 
here the first complete result for a 
two-loop three point Green's function in 
the MSSM.    

The $gg \to h$ amplitude  at 
${\cal O} \left(\alpha_s^3\right)$ includes
135 two-loop Feynman diagrams. 
We have generated them using 
QGRAF~\cite{qgraf}. 
We implemented the MSSM Feynman rules 
following the method of Ref.~\cite{majorana1} for Majorana 
fermions. Traces of Dirac matrices and the color algebra are carried out 
with programs written in standard algebraic manipulation packages, 
including FORM~\cite{form}. We have checked that our computer 
programs and  FeynArts~\cite{feynarts} generate equivalent integrands 
for the amplitude.  
\newcommand{\strs}{0mm}
\newcommand{\dwith}{0.12 \textwidth}
\newcommand{\abstand}{\rule{\strs}{0.09\textwidth}} 
\newcommand{\vs}{\rule{0.015\textwidth}{\strs}}
\begin{figure}[t]
\begin{tabular}{ccc}
\begin{minipage}[c]{\dwith}
\includegraphics[width=\textwidth]{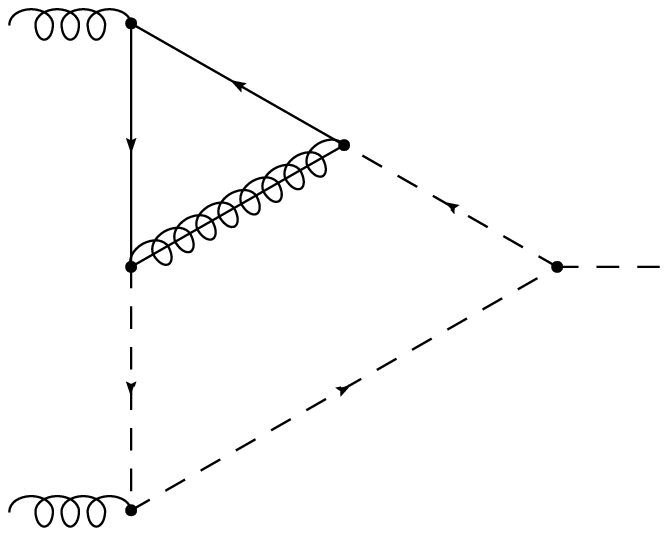}
\end{minipage} \vs
&
\begin{minipage}[c]{\dwith}
\includegraphics[width=\textwidth]{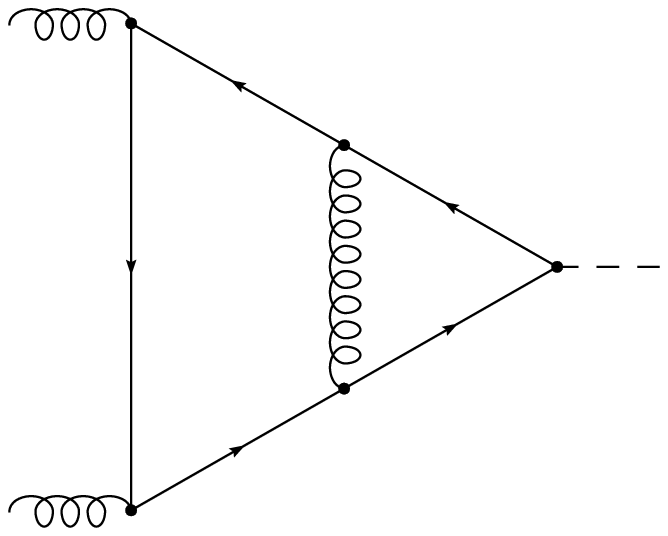}
\end{minipage} \vs
&
\begin{minipage}[c]{\dwith}
\includegraphics[width=\textwidth]{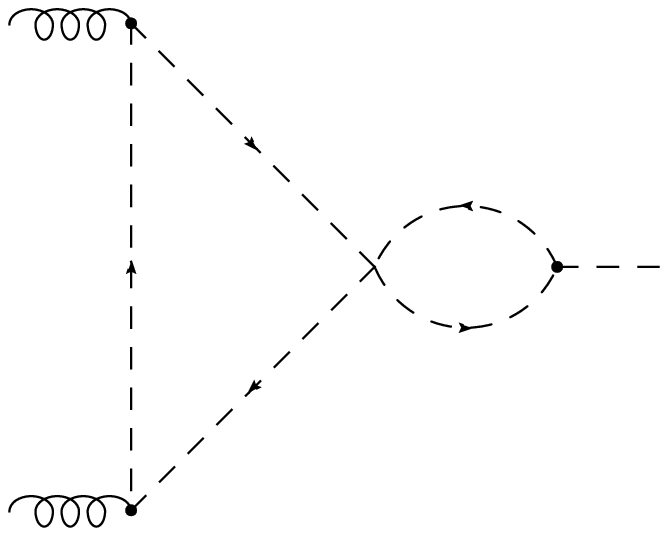}
\end{minipage}
\\
\abstand
\begin{minipage}[c]{\dwith}
\includegraphics[width=\textwidth]{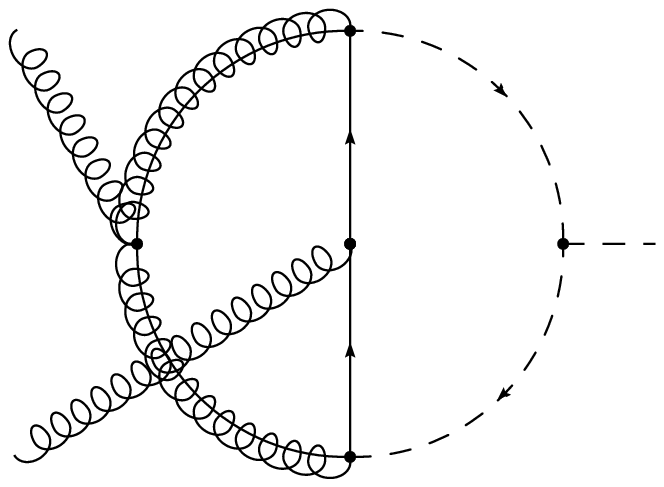}
\end{minipage} \vs
&
\begin{minipage}[c]{\dwith}
\includegraphics[width=\textwidth]{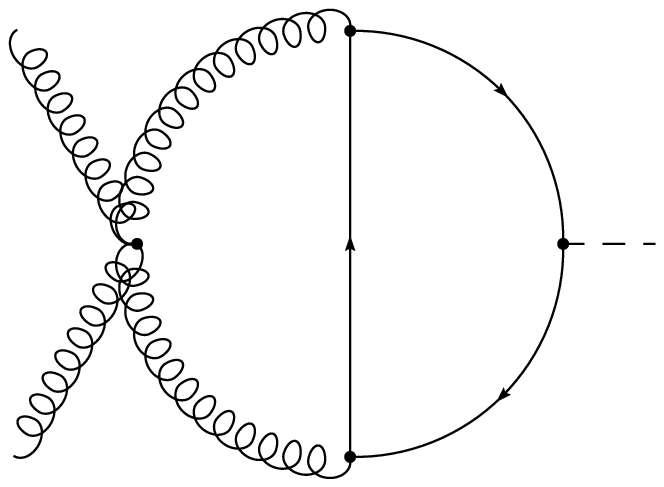}
\end{minipage} \vs
&
\begin{minipage}[c]{\dwith}
\includegraphics[width=\textwidth]{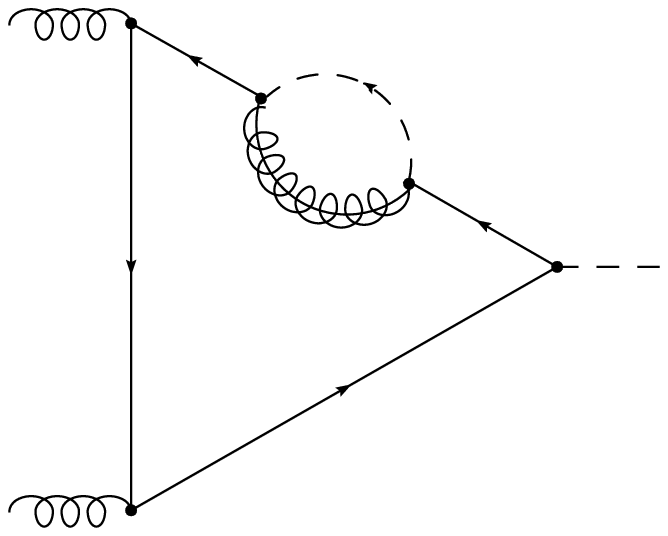}
\end{minipage}
\end{tabular}
\caption{Sample of two-loop diagrams containing up to four different 
mass parameters in the propagators.
\label{fig:diagrams}}
\end{figure}

We depict some of the Feynman diagrams in Fig.~\ref{fig:diagrams}. We  
computed diagrams with only one massive particle in the loops both 
analytically~\cite{bucherer} and numerically~\cite{beerli}. We computed 
the remaining diagrams numerically applying the method 
of~\cite{beerli,lazopoulos}. 
We review here only the salient features of the numerical method.  

We  combine propagators in each loop with 
a set of Feynman parameters and integrate out 
the loop momenta. The Feynman parameters are mapped to a unit hypercube 
integration domain.  
At a second step, we apply a sector decomposition 
algorithm~\cite{sector1,sector3} 
to factorize divergences in the Feynman parameters 
which are regulated by the  dimensional parameter $D=4-2\epsilon$. 
We deform the contour of integration away from the poles which 
may occur for physical values of the external momenta and mass 
parameters~\cite{cdeform1}.  
Finally, we extract the factorized singularities in $\epsilon$ with 
independent subtractions on the real part of Feynman parameters and 
perform an expansion around $\epsilon =0$~\cite{beerli}.  
With this procedure the 
amplitude is written in terms of ${\cal O}(1000)$ integrals 
which can be evaluated with standard numerical methods. We have used 
the Divonne and Cuhre integration algorithms from the Cuba 
library~\cite{cuba}.

When we apply this method naively to 
diagrams  such as the last diagram 
in Fig.~\ref{fig:diagrams} with a sbottom-gluino bubble  
inserted in a bottom propagator, the numerical integration shows 
a very poor convergence. 
Our technique is based on a deformation of Feynman parameters: 
\[ x_i \to z_i = x_i + i \lambda_i y_i \]
The function $y_i$ is constructed using the prescription of 
~\cite{cdeform1,beerli}, and it depends 
on Feynman parameters and the mass parameters. 
We can usually set a common numerical value for the constants 
$\lambda_i$, which dictate the magnitude 
of the deformation, when  all mass values 
are of the same order of magnitude. 
However, the large hierarchy of the squared  mass values 
for the bottom quark and the gluino or the bottom squarks may 
yield $y_i$'s of disparate magnitude for different 
Feynman parameters.  Small values of $y_i$ may not  distance the 
contour of deformation sufficiently from singularities. In turn, 
large  $y_i$ values may result to a contour which is not equivalent 
with the integration over the original Feynman parameters.   
We can remedy these problems  by adjusting the 
parameters $\lambda_i$ independently for each Feynman 
parameter.  We perform a first Monte-Carlo integration 
with a small number of integrand evaluations 
setting all $\lambda_i$ to a common value $\lambda$. We use this 
integration in order to obtain an estimate of the maximum values 
$y_i^{max}$ of $y_i$.  
We then  compute the  integrals setting $\lambda_i \sim \lambda /y_i^{max}$, 
where  $\lambda$ has  a typical value $0.1 - 0.5$. 
With this refined selection of the parameters which determine the 
deformation  of the integration contour we are able to evaluate all 
diagrams efficiently. 

We were also able to compute the problematic diagrams with a uniform 
selection of $\lambda_i$ values by using a Feynman parameterization 
similar to the one in Ref.~\cite{sector3}. 
This parameterization casts diagrams with bubble subgraphs 
as a sum of two terms.  One of  them 
corresponds to  an one-loop integral and matches to the 
counter-term for mass renormalization. The second term corresponds to 
a two-loop integral which can be evaluated without a specially tuned 
contour deformation.

We have used conventional dimensional 
regularization (DREG)~\cite{dreg}, dimensional 
reduction (DRED)~\cite{dred1,dred2} and the 
four dimensional helicity scheme (FDH)~\cite{fdh} 
with minimal subtraction. 
In all schemes an anti-commuting 
$\gamma_5$ prescription was employed. 
Similarly to~\cite{mssmeff1}, we decouple  the top-quark, squarks and 
gluino from the running of the strong coupling $\alpha_s$. In addition, 
we perform a pole renormalization of all masses and renormalize the
squark higgs couplings as in~\cite{Bartl:1997yd}. 

DREG is known to violate the symmetries of the MSSM. DRED 
is a consistent renormalization scheme for the computation of the 
MSSM amplitude through the next to leading order in the 
strong coupling. The two schemes are, however,
related by finite shifts in the 
coupling constants and masses~\cite{Jack:1994bn}. This equivalence 
only holds when new terms are added to the Lagrangian in 
DRED~\cite{Jack:1993ws,Jack:1994bn,Jack:1994kd}: operators
with $\epsilon$-scalars which
arise when the $4$-dimensional 
gluon field is split into a $D$-dimensional part and
its remaining $4-D$ components. Notably, in the MSSM, the only relevant
term for the amplitude $gg\rightarrow h$ is the mass 
term for the
$\epsilon$-scalars. As this mass can be always absorbed into
a redefinition of the squark masses~\cite{Jack:1994rk}, the calculation in the
MSSM can be performed by setting it to zero. 

The picture described above changes dramatically when considering a theory 
with less symmetries than the MSSM, where 
supersymmetry is only softly broken  and the possible 
$\epsilon$-scalar couplings are still very restricted.
We observed that, if either $SU(2)_L$ 
or  (softly broken) supersymmetry are absent, a new coupling between 
the $\epsilon$-scalars and the higgs fields emerges radiatively. 
This coupling is indispensable in order to render 
the one-loop cross-section for the process 
$h \to \epsilon \epsilon$ to be of ${\cal O}(\epsilon)$ (evanescent).  
In the MSSM this happens automatically 
due to a cancellation among contributions 
from up and down type quarks and the corresponding squarks. 
In the SM this cancellation does not take place. 
Most importantly, the $\epsilon$-higgs coupling 
and its consistent renormalization 
should be included in order for the results  for the 
SM two-loop amplitude $gg \to h$ in DRED and DREG to agree. 

We compared the results in the two schemes  
after we  accounted for known shifts in the  strong 
coupling $\alpha_s$ and the mass parameters between the two schemes.    
We  found that the two results agree only if an additional shift in 
the higgs-squark-squark coupling is performed.  
We find that the relation between the 
renormalized coupling $m_q^2 V_{h\tilde q_i \tilde q_j}$ in DREG and 
DRED is,
\begin{eqnarray}
 \label{eq:shift}
&&\left( m_q^2 V_{h\tilde q_i \tilde q_j} \right)^{{\rm DREG}} - 
\left( m_q^2 V_{h\tilde q_i \tilde q_j} \right)^{{\rm DRED}}\nonumber\\
&&\quad=\frac{\partial \left(m_q^2 V_{h\tilde q_i \tilde q_j}\right)}{\partial m_q}
\left[ 
\delta^{{\rm DRED}}_{m_q} - \delta^{{\rm DREG}}_{m_q}
\right] 
+ {\cal O} \left( \alpha_s^2\right),
\end{eqnarray}
where $V_{h\tilde q_i \tilde q_j}$ is the dimensionless part of the tree 
coupling, and $\delta^{{\rm DRED}}_{m_q}$,$\delta^{{\rm DREG}}_{m_q}$ are  
the pole  mass-renormalization counter-terms in the two schemes. 
We have verified that the shift in Eq.~\ref{eq:shift} 
is also needed for the computations of the decay rate
$h \to \tilde q_i \tilde q_j$ to agree in DRED and DREG at one loop. The reason
for this shift is simple. The one loop corrections to the decay amplitude 
$h \to \tilde q_i \tilde q_j$ are identical in DRED and DREG 
before renormalization. 
However, the renormalization scheme for the squark higgs coupling 
in~\cite{Bartl:1997yd} involves the quark pole mass, which is related to the 
bare mass trough  a scheme dependent expression. 
The shift in Eq.~\ref{eq:shift} simply cancels that dependence in such a 
way that the DRED result for the decay rate is recovered. 

We can avoid to compute  the contribution of operators with 
$\epsilon$-scalars to the $gg \to h$ amplitude for  the MSSM 
by using  the FDH scheme. 
DRED and FDH treat external polarizations differently; this gives rise 
to a  relative factor of  $(1-\epsilon)^2$  in the two-schemes 
for the squared amplitudes through ${\cal O}(\alpha_s^3)$. Diagrams with 
internal $\epsilon$ scalars in DRED are accounted for in the FDH scheme 
by diagrams with internal $D_S$-dimensional gluons, where $D_S$ is the 
dimension of the spin algebra. The dimensionality of the loop integrals 
$D$ is kept distinct with $D < D_S$;  after performing the loop-integrations
an analytic continuation of $D_s$ to 4 dimensions~\cite{fdh} takes place. 
The FDH scheme cannot account for the contributions of diagrams with the 
$\epsilon$-higgs coupling. As we discussed, these are needed
for the  $gg \to h$ two-loop amplitude in less symmetric theories than the 
MSSM, such as the SM. The discrepancy cannot  be absorbed 
in any coupling or mass redefinition, and the FDH result is inconsistent 
with the results in DRED and DREG for the SM two-loop amplitude.

\begin{figure}[th]
\begin{center}
\includegraphics[width=0.5\textwidth]{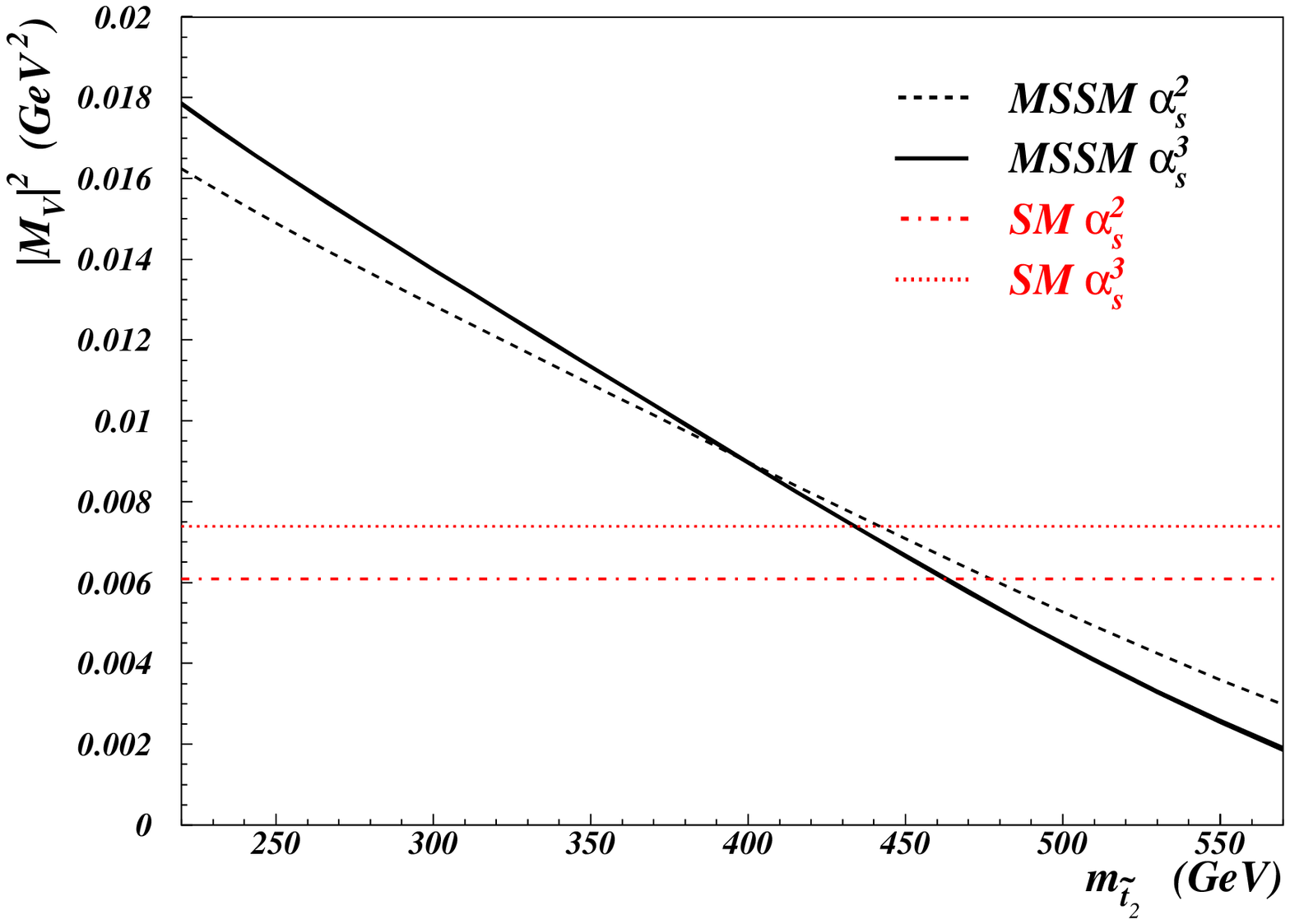}
\includegraphics[width=0.5\textwidth]{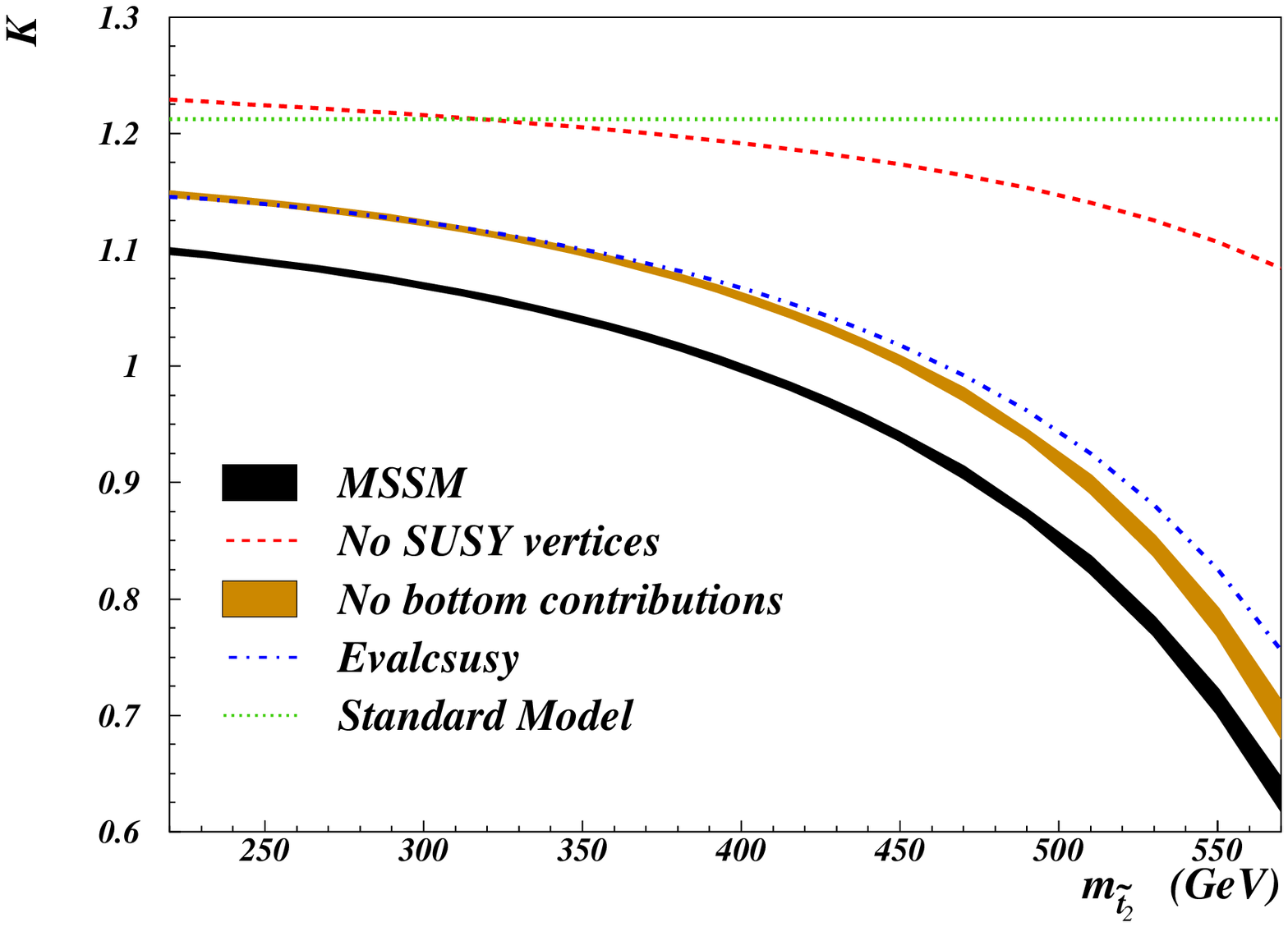}
\caption{The squared UV and IR renormalized amplitude for a light Higgs 
boson through ${\cal O} (\alpha_s^3)$ and the corresponding K-factors}
\label{fig:light_top}
\end{center}
\end{figure}
    
We now present numerical results for the MSSM two-loop 
amplitudes $gg \to h,H$.  We neglect here the Higgs couplings 
to quarks and squarks other than the ones in the third generation. 
The two-loop amplitudes are infrared divergent with  
poles up to second order in the dimension parameter $\epsilon$. 
The singular part is universal~\cite{IRuni1} and cancels against 
other universal contributions at the same order in $\alpha_s$ 
from real radiation processes.  We present here the finite part after UV 
renormalization in the $\overline{{\rm MS}}$ scheme and subtracting  
the infrared counter-term of Ref.~\cite{IRcatani}. 
A complete phenomenological analysis  
requires the inclusion of the non-singular parts from real radiation 
and will be the subject 
of a future publication. Nevertheless, useful conclusions may also 
be inferred from solely the two-loop amplitudes since 
they include all  diagrams with more than one massive internal particle. 
These are the diagrams which had not been computed 
earlier in the literature~\cite{bucherer,bonciani1,spira-squarks,bonciani2} 
and they form a subset which is infrared finite. 

In Figure~\ref{fig:light_top} we present our results for the production of a 
light higgs in gluon fusion in the MSSM.
On the upper panel we show the renormalized squared amplitude for a light 
neutral Higgs boson $gg \to h$ through order ${\cal O}(\alpha_s^3)$, 
averaging over external gluon polarizations and colors. 
As discussed above, we subtracted a universal infrared 
counter-term in order to
obtain a finite result. We show the SM value as a reference. 
The lower panel shows the corresponding K-factor: the 
ratio of the  squared amplitude
through ${\cal O}(\alpha_s^3)$ divided by the ${\cal O}(\alpha_s^2)$ result. 
We   also include the K-factors obtained in various approximations of our
full result. We write our results in terms of the 
$\alpha_s^{\overline{{\rm MS}}}$ with 
$\alpha_s^{\overline{{\rm MS}}}(M_Z)=0.1176$.
We choose a relatively small mass value for one of the stop 
squarks 
$m_{{\tilde t}_1} = 150 \,{\rm GeV}$ and vary the mass of the heavy stop quark
$m_{{\tilde t}_2}$.  In the bottom sector we have set $m_b=5 \,{\rm GeV}$,
$m_{{\tilde b}_1}=350\,{\rm GeV}$ and $m_{{\tilde b}_2}=370\,{\rm GeV}$. 
For the SUSY parameters, we have chosen $\alpha=3^{\circ}$, $\tan \beta = 20$, 
$\mu=300\,{\rm GeV}$, and the squark mixing angles 
$\theta_{\tilde{t}}=\theta_{\tilde{b}}=40^{\circ}$. 
Finally, $m_{\rm gluino} = 500 \,{\rm GeV}$, $m_{h} = 115 \,{\rm GeV}$, 
the renormalization scale is fixed to the higgs mass, 
$\mu_{\rm ren}= m_h$. The renormalization of the squark mixing 
angles is done as in~\cite{mssmeff1} at a  
scale  $\mu_{{\theta}}=200\,{\rm GeV}$. 

The squared amplitude in the MSSM decreases for growing 
$m_{\tilde{t}_2}$. This is due to large cancellations between 
diagrams  where  a top or a  light stop couple to the Higgs boson. 
The ${\cal O}(\alpha_s^3)$ contribution ranges from $15 \%$ to $-40\%$, 
becoming negative at large  values of $m_{\tilde{t}_2}$. 
A significant part of this correction originates 
from the infrared finite subset  of diagrams with gluinos and 
squark quartic couplings; their contribution to the squared amplitude is 
negative and grows in absolute value with growing 
$m_{\tilde{t}_2}$. The contribution from bottom and sbottom loops
is below $3\%$ for this value of $\tan\beta$, and smaller for lower
values of $\tan\beta$. We also show the K-factor 
obtained with  the effective theory calculation of 
Ref.~\cite{mssmeff1} using the published program {\tt evalcsusy}. 
The approximation given by this calculation is remarkably 
good. At small values of $m_{\tilde{t}_2}$ it almost coincides with the MSSM
result when neglecting the bottom contributions whereas at larger values of
the mass splitting it differs only by a few percent. The growing
(small) discrepancy is due to the contribution of the heavy stop decreasing
with a larger mass $m_{\tilde{t}_2}$ rendering  the  contribution of 
light stop loops, 
which are not perfectly approximated by the effective theory, more 
significant.

The scenaria with a light scalar quark have received recent attention. 
In particular, a large mass gap in the two scalar-top physical states 
reduces the ``fine-tuning''of the MSSM~\cite{perelstein}; 
this may be an outcome of supersymmetry breaking via ``mirage'' 
mediation~\cite{cho}. In Fig~\ref{fig:light_top} we observe that  
the squared amplitude decreases as the squark mass splitting increases. 
The parameter region where both stop squarks are light is 
excluded by measurements of  the 
$\rho-$parameter~\cite{perelstein}. However, as the  stop 
mass difference increases we obtain viable parameter values; they 
result to a $ggh$ interaction which is significantly weaker 
than in the SM; the discovery of a light Higgs boson may then become 
very difficult at the LHC.

\begin{figure}[th]
\begin{center}
\includegraphics[width=0.5\textwidth]{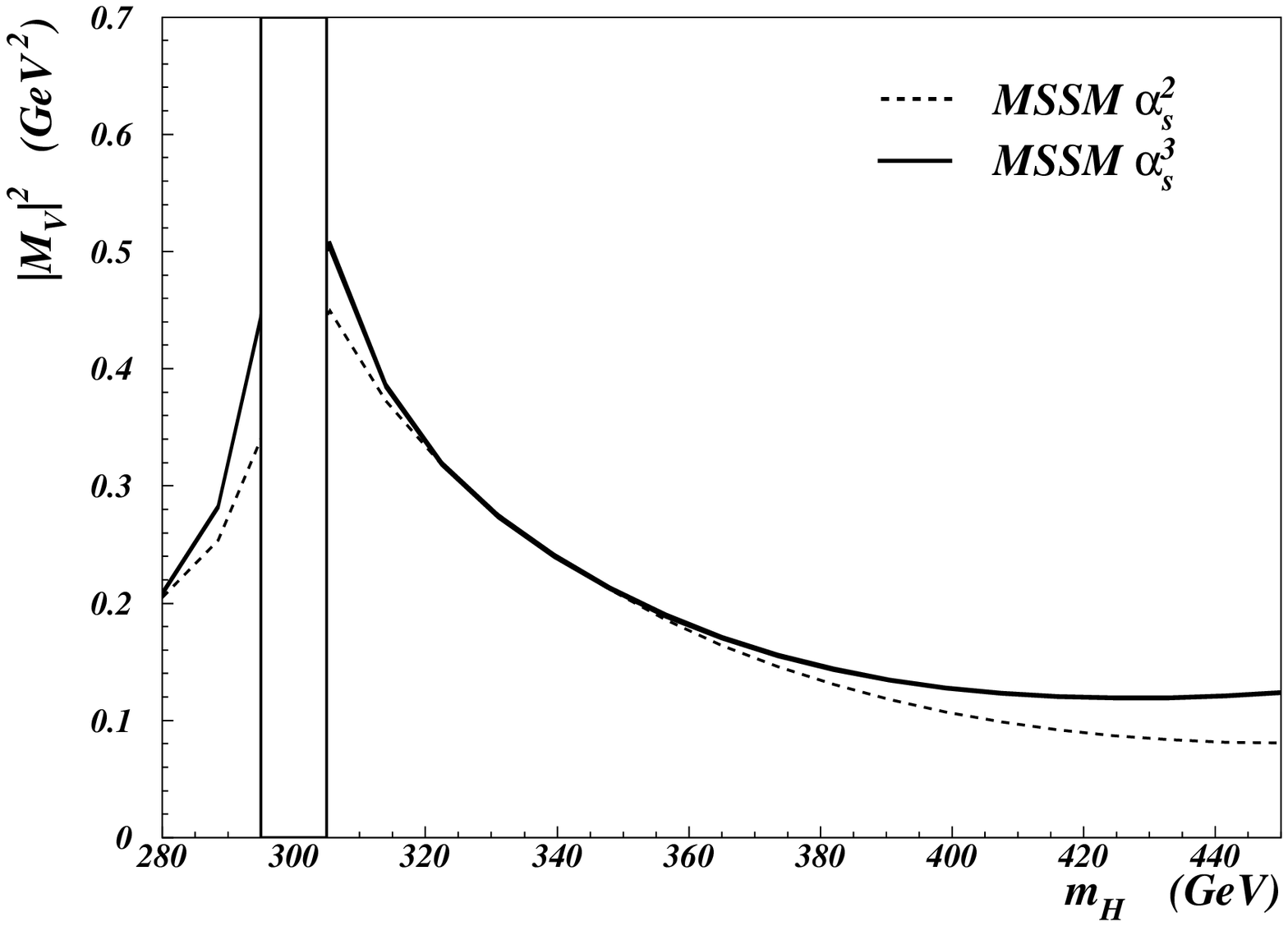}
\includegraphics[width=0.5\textwidth]{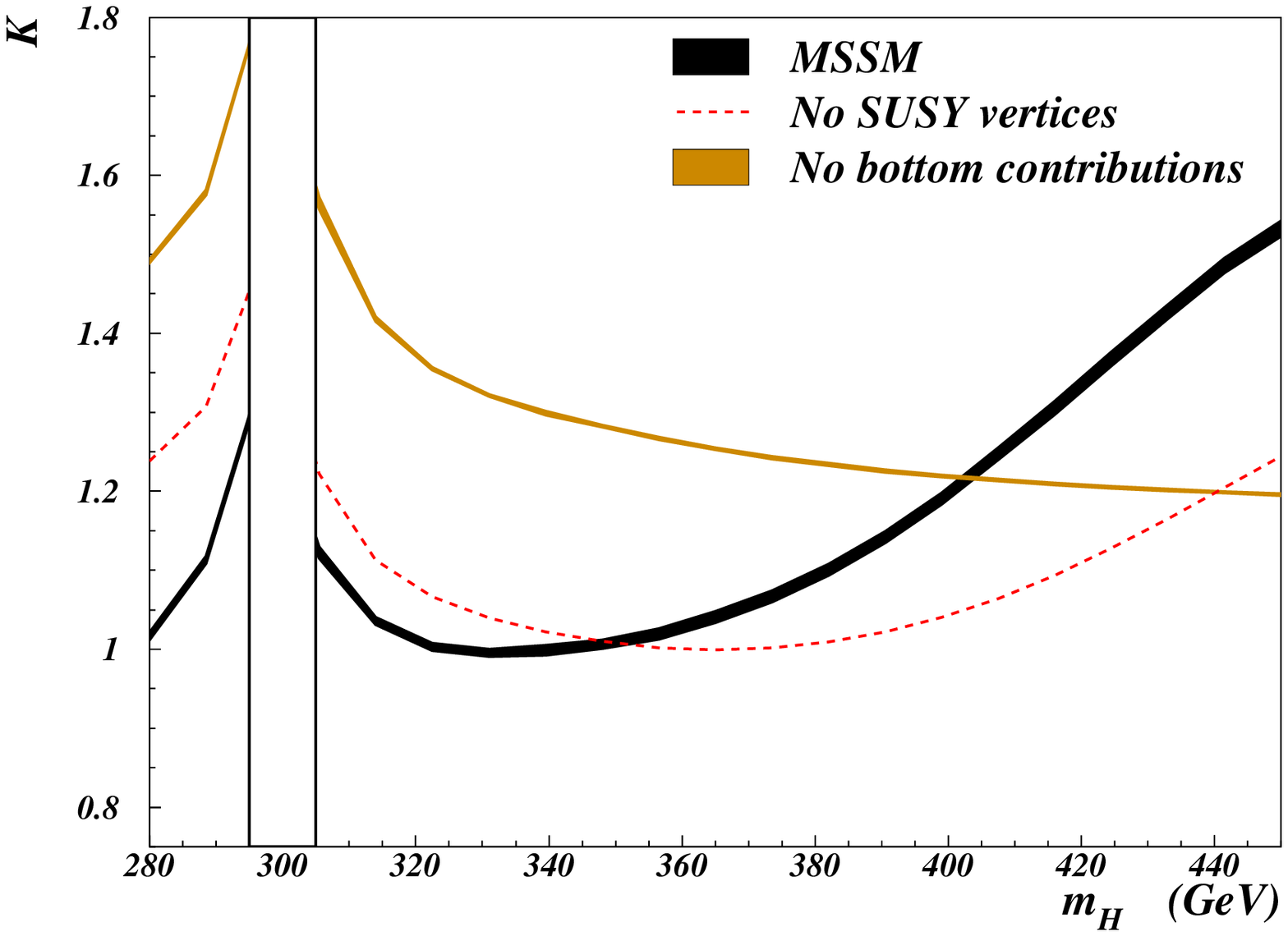}
\caption{The squared UV and IR renormalized amplitude for a heavy neutral 
Higgs boson through ${\cal O} (\alpha_s^3)$ and the corresponding K-factors}
\label{fig:heavy_top}
\end{center}
\end{figure}

In Figure~\ref{fig:heavy_top} we show the corresponding results for the
production of a heavy Higgs. In this case we fixed the heavy stop mass
$m_{\tilde{t}_2}=350\,{\rm GeV}$ and varied the heavy Higgs boson mass. We 
set the renormalization scale to $\mu_{\rm ren} = m_H$. 
All the other masses and parameters are identical to the light Higgs boson 
calculation described above. 
At $m_H=2\,m_{\tilde{t}_1}=300\,{\rm GeV}$ there is a threshold,
where the perturbative calculation diverges. In the plots, we have 
conservatively whitewashed a window of $5\,{\rm GeV}$ around the 
threshold, but we have checked that 
our numerical calculation works fine for phase space points much closer to 
this threshold. We find that the ${\cal O}(\alpha_s^3)$ corrections
in the MSSM are very mild, growing to about $20\%$ in the region of the 
threshold but ammounting to only a few percent anywhere else. 
The bottom sector contributions are, however, very important, 
ammounting to almost $40\%$ for small Higgs boson masses. 
As in the light Higgs boson case, contributions from the diagrams with 
gluinos and squark quartic couplings are substantial.

In this paper we have computed the full two-loop amplitudes  
$gg \to h$ and $gg \to H$ in the MSSM, a complicated  extension 
of the Standard Model. We have 
developed a powerful new method for multi-loop  
calculations and  applied it to compute 
the first two-loop three-point amplitude known in the MSSM 
for arbitrary masses of sparticles. 
Our results will improve the precision 
of cross-section predictions for the gluon-fusion process in 
the MSSM. We are looking forward to further applications of  our method.

{\bf Acknowledgements:} 
We thank Zvi Bern, Giuseppe Degrassi, Lance Dixon, Uli Haisch, 
Robert Harlander, David Kosower,  Zoltan Kunszt, Kirill Melnikov, 
Pietro Slavich, Matthias Steinhauser and James Wells for useful discussions.  
This research was supported by the Swiss National Science Foundation  
under contracts 200021-117873 and 200020-113567/1.

\end{document}